\documentclass[conference,9pt]{IEEEtran}
\IEEEoverridecommandlockouts
\usepackage{cite}
\usepackage{amsmath,amssymb,amsfonts}
\usepackage{algorithmic}
\usepackage{graphicx}
\usepackage{textcomp}
\usepackage{xcolor}
\usepackage{multirow}
\usepackage{svg}
\usepackage{booktabs}
\usepackage{threeparttable}
\usepackage{hyperref}
\def\BibTeX{{\rm B\kern-.05em{\sc i\kern-.025em b}\kern-.08em
    T\kern-.1667em\lower.7ex\hbox{E}\kern-.125emX}}

\newcommand{\squishlist}{
 \begin{list}{$\bullet$}
  { \setlength{\itemsep}{0pt}
     \setlength{\parsep}{3pt}
     \setlength{\topsep}{3pt}
     \setlength{\partopsep}{0pt}
     \setlength{\leftmargin}{1.5em}
     \setlength{\labelwidth}{1em}
     \setlength{\labelsep}{0.5em} } }
\newcommand{\squishend}{
  \end{list}  }

\newcommand{\papername}{DiffuSE}

\newcommand{\ppaimprove}{$147\%$}
\newcommand{\hviimprove}{$96.6\%$}

\begin{document}

% \title{Conference Paper Title*\\
% {\footnotesize \textsuperscript{*}Note: Sub-titles are not captured for https://ieeexplore.ieee.org  and
% should not be used}
% \thanks{Identify applicable funding agency here. If none, delete this.}
% }
% \title{DeepDataflow: Leveraging Diffusion Models for Cross-Layer Optimization of Spatial Architecture
% }
\title{\papername: Cross-Layer Design Space Exploration of DNN Accelerator via Diffusion-Driven Optimization}

% \author{
% \IEEEauthorblockN{
% Yi Ren\textsuperscript{2},
% Chenhao Xue\textsuperscript{1},
% Jiaxing Zhang\textsuperscript{1},
% Chen Zhang\textsuperscript{3},
% Qiang Xu,
% Yibo Lin\textsuperscript{1},
% Lining Zhang,
% Guangyu Sun\textsuperscript{1}
% }
% \IEEEauthorblockA{\textsuperscript{1}\textit{School of Integrated Circuits, Peking University}, Beijing, China}
% \IEEEauthorblockA{\textsuperscript{2}\textit{School of Software and Microelectronics, Peking University}, Beijing, China}
% \IEEEauthorblockA{\textsuperscript{3}\textit{School of Electronics, Information and Electrical Engineering, Shanghai Jiao Tong University}, Shanghai, China}
% }
\author{
    \IEEEauthorblockN{
        Yi Ren\textsuperscript{1,2}, 
        Chenhao Xue\textsuperscript{1},
        Jiaxing Zhang\textsuperscript{1}, 
        Chen Zhang\textsuperscript{3},
        Qiang Xu\textsuperscript{4,8},
        Yibo Lin\textsuperscript{1,5,6},
        Lining Zhang\textsuperscript{7},
        Guangyu Sun\textsuperscript{1,5,6,*}
        \thanks{\textsuperscript{*}Corresponding author: Guangyu Sun (\href{mailto:gsun@pku.edu.cn}{gsun@pku.edu.cn}).}
    }
    \IEEEauthorblockA{
        \textsuperscript{1}\textit{School of Integrated Circuits}, \textsuperscript{2}\textit{School of Software and Microelectronics, Peking University}, Beijing, China \\
         % Beijing, China \\
        \textsuperscript{3}\textit{Shanghai Jiao Tong University}, Shanghai, China \\
        \textsuperscript{4}\textit{Department of Computer Science and Engineering, The Chinese University of Hong Kong}, Sha Tin, Hong Kong S.A.R. \\
        \textsuperscript{5}\textit{Institute of Electronic Design Automation, Peking University}, Wuxi, China \\
        \textsuperscript{6}\textit{Beijing Advanced Innovation Center for Integrated Circuits}, Beijing, China \\
        \textsuperscript{7}\textit{School of Electronic and Computer Engineering, Peking University}, Shenzhen, China \\
        \textsuperscript{8}\textit{National Center of Technology Innovation for EDA}, Nanjing, China \\
        % \{xch927027,gsun\}@pku.edu.cn
    }
    \{yiren20, zjx\}@stu.pku.edu.cn, \{xch927027, yibolin, eelnzhang, gsun\}@pku.edu.cn, chenzhang.sjtu@sjtu.edu.cn, qxu@cse.cuhk.edu.hk
}

\maketitle

\begin{abstract}
    The proliferation of deep learning accelerators calls for efficient and cost-effective hardware design solutions, where parameterized modular hardware generator and electronic design automation (EDA) tools play crucial roles in improving productivity and final Quality-of-Results (QoR).
    To strike a good balance across multiple QoR of interest (e.g., performance, power, and area), the designers need to navigate a vast design space, encompassing tunable parameters for both hardware generator and EDA synthesis tools.
    However, the significant time for EDA tool invocations and complex interplay among numerous design parameters make this task extremely challenging, even for experienced designers.
    To address these challenges, we introduce \papername{}, a diffusion-driven design space exploration framework for cross-layer optimization of DNN accelerators.
    \papername{} leverages conditional diffusion models to capture the inverse, one-to-many mapping from QoR objectives to parameter combinations, allowing for targeted exploration within promising regions of the design space.
    By carefully selecting the conditioning QoR values, the framework facilitates an effective trade-off among multiple QoR metrics in a sample-efficient manner.
    Experimental results under 7nm technology demonstrate the superiority of the proposed framework compared to previous arts.
\end{abstract}

% By leveraging conditional diffusion models to capture the inverse mapping from QoR objective space to parameter design space, \papername can efficiently identify potential parameter combination with desired QoR metrics.
% In addition, a customized acquisition function is adopted to select appropriate conditioning objectives, which closely adheres to known prior while maximizing the Expected HyperVolume Improvement (EHVI) during multi-objective optimization. 

\begin{IEEEkeywords}
diffusion models, design space exploration, cross-layer optimization.
\end{IEEEkeywords}

\section{introduction}
\label{sec:introduction}

% Introduce DNN accelerator, automatic generation framework, and ASIC implementation
% Domain-specific accelerators (DSAs), particularly those tailored for deep neural networks (DNNs), are increasingly deployed from datacenters to edge devices.
% This growing trend emphasizes the need for productive and cost-efficient hardware design, fueling research into highly-parameterized and modular hardware generators ~\cite{wei2017automated,cong2018polysa,venkatesan2019magnet,genc2021gemmini,jia2021tensorlib,luo2023rubick}.
% For varying target scenarios and design objectives of ASIC accelerators, these hardware generators can swiftly produce synthesizable RTL implementations.
% The RTL designs are further processed by electronic design automation (EDA) tools for logic synthesis and physical design, eventually yielding a manufacturable layout.

Domain-specific accelerators (DSAs), especially those designed for deep neural networks (DNNs), are increasingly deployed across diverse platforms, from datacenters to edge devices, highlighting their critical role in enabling efficient AI computation. 
This growing importance drives the demand for productive and cost-effective hardware design methodologies, spurring research into highly parameterized and modular hardware generators capable of rapidly producing synthesizable RTL implementations~\cite{wei2017automated,cong2018polysa,venkatesan2019magnet,genc2021gemmini,jia2021tensorlib,luo2023rubick}. 
The generated RTL designs are processed through electronic design automation (EDA) tools for logic synthesis and physical design, ultimately yielding manufacturable layouts. 
However, achieving high Quality-of-Results (QoR) in DNN accelerator designs is influenced by a myriad of factors. 
Hardware design parameters, such as multiply-accumulate (MAC) array size, interconnection style, and on-chip buffer size, directly affect performance, power consumption, and area footprint (PPA). 
Similarly, EDA tool parameters, including target delay, synthesis effort, and optimization heuristics, play a crucial role in determining PPA and meeting design constraints. 
Together, these parameters create an enormous and complex space, making the design space exploration (DSE) of optimal configurations exceedingly challenging, even for experienced engineers, given the multi-staged and computationally intensive VLSI design flow.

\begin{figure}[!t]
    \centering
    \includegraphics[width=\linewidth]{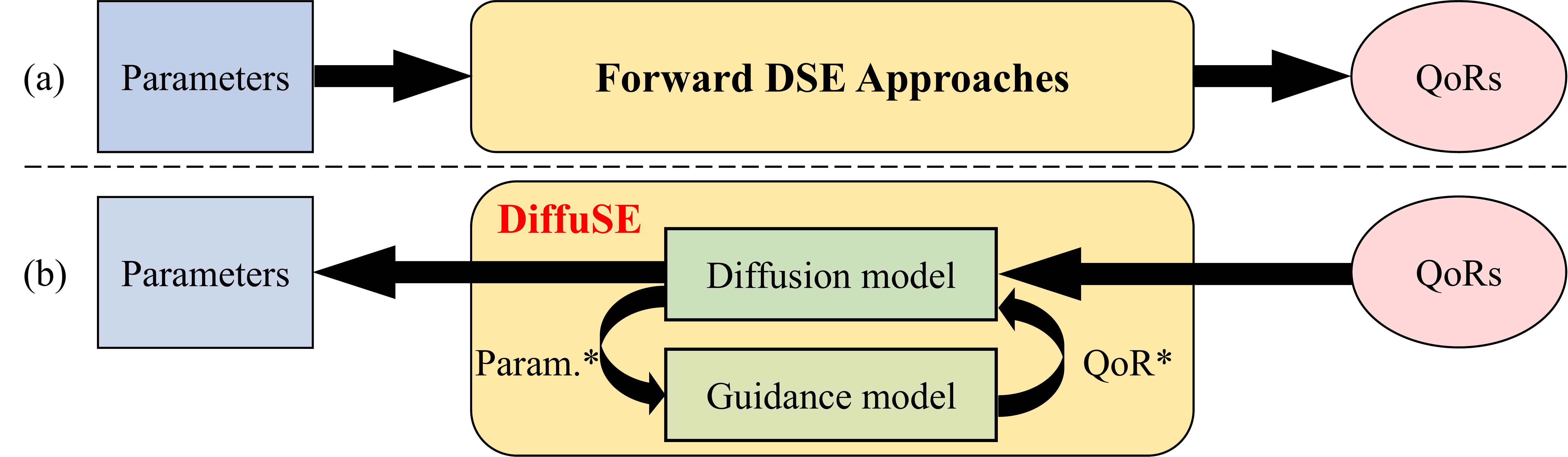}
    \caption{Comparison of the forward design space exploration (DSE) approaches, and the inverse DSE approach proposed by \papername{}.}
    \label{fig:comparison}
\vspace{-10pt}
\end{figure}

Significant progress has been made in addressing the challenge of DSE for physically optimal designs. 
Tools such as BOOM-explorer~\cite{bai2021boom} and SoC-Tuner~\cite{chen2024soctuner} explore the architectural design space while running complete VLSI flows to evaluate QoR. 
At the EDA level, machine learning-based approaches have been developed to automate the tuning of tool parameters~\cite{kwon2019learning,xie2020fist,liang2021flowtuner,geng2022ptpt,geng2022ppatuner}. 
Recently, cross-layer DSE methods have emerged, enabling joint optimization of PPA across design and EDA layers, as demonstrated in works on adders~\cite{ma2018cross,geng2021high}.
These DSE methods share a \textbf{forward approach}, where parameter-to-QoR prediction models are leveraged to reduce the reliance on extensive VLSI evaluations, and thereby enhance the exploration efficiency, as shown in Fig.~\ref{fig:comparison}(a).
Among all the forward approaches, one of the most prominent examples is Bayesian Optimization~\cite{bai2021boom,chen2024soctuner,geng2022ppatuner,ma2018cross,geng2021high,geng2022ptpt}.

Nevertheless, the forward approach faces \textbf{generalization issue}: the prediction models typically require substantial training data for accurate estimatation of QoRs. Yet, in the context of DNN accelerator DSE, obtaining training data with extensive VLSI evaluations can be prohibitively costly. Consequently, for parameter configurations that deviate from the training distribution, the prediction models suffer from significant accuracy degradation, which may hinder DSE efficiency.

To address the challenge of generalization, our key insight is to predict the QoR of parameter configurations close to the training dataset. 
Based on this, we propose \papername{}, a novel cross-layer DSE framework that leverages diffusion models to generate in-distribution parameter configurations (see Fig.~\ref{fig:comparison}(b)). 
Diffusion models have demonstrated their capability to generate complex data structures such as images~\cite{sohl2015deep,ho2020denoising}, text~\cite{li2022diffusion}, and speech~\cite{kim2022guided}. 
Similarly, by representing configurations as structured data, the diffusion model learns the dataset distribution and generates configurations with similar characteristics. 
Moreover, with appropriate guidance such as classification labels~\cite{ho2022classifier} or text prompts~\cite{kim2022guided}, diffusion models can \textit{conditionally} generate diverse high-quality samples that match the guidance.
Building on this, \papername{} integrates a guidance module to steer the diffusion model toward sampling configurations within subspaces aligned with target QoR objectives. 
Leveraging the diffusion module, the guidance module operates within the dataset distribution, avoiding generalization issues typically caused by out-of-distribution sampling.
Together, these components enable the model to learn the mapping from QoR metrics to feasible parameter configurations and directly sample promising configurations, making it an \textbf{inverse approach}.
% By carefully selecting conditioning QoR values with the proposed Pareto-aware conditioning mechanism, \papername{} consistently explores high-potential design configurations, optimizing for multiple target QoR metrics.
% The experiment results demonstrates that \papername{} can find optimized combination of hardware parameters and EDA tool parameters, with improved sample efficiency over previous arts.
By employing a Pareto-aware conditioning mechanism to select target QoR values, \papername{} effectively explores high-potential design configurations and optimizes multiple QoR metrics simultaneously. 
Experimental results demonstrate that \papername{} achieves superior sampling efficiency compared to prior methods, identifying optimized combinations of hardware and EDA tool parameters.

The main contribution of this work is summarized as follows:
\begin{itemize}
    \item We propose \papername, a comprehensive framework for jointly optimizing hardware architecture and synthesis configurations to design optimized DNN accelerators.
    \item We utilize diffusion models to capture the complex inverse mapping from objective PPA space to design parameter space.
    \item We propose a heuristic mechanism to choose proper conditioning objective values at each iteration.
    \item We evaluate \papername{} against a widely used approach, and demonstrate that \papername{} can improve the PPA by \ppaimprove{} and hypervolume by \hviimprove{}, with superior efficiency over the previous method.
\end{itemize}

The remainder of this paper is organized as follows:
Section~\ref{sec:background} provides preliminary on diffusion models and the problem formulation.
Section~\ref{sec:methodology} details the \papername{} optimization framework.
Section~\ref{sec:experiment} presents the experimental results.
Finally, Section~\ref{sec:conclusion} concludes the paper.

\section{background}
\label{sec:background}

In this section, we provide our problem formulation and some background knowledge about conditional diffusion models in our context for a better understanding.

\subsection{DNN Accelerator}

\begin{figure}[!t]
    \centering
    \includegraphics[width=\linewidth]{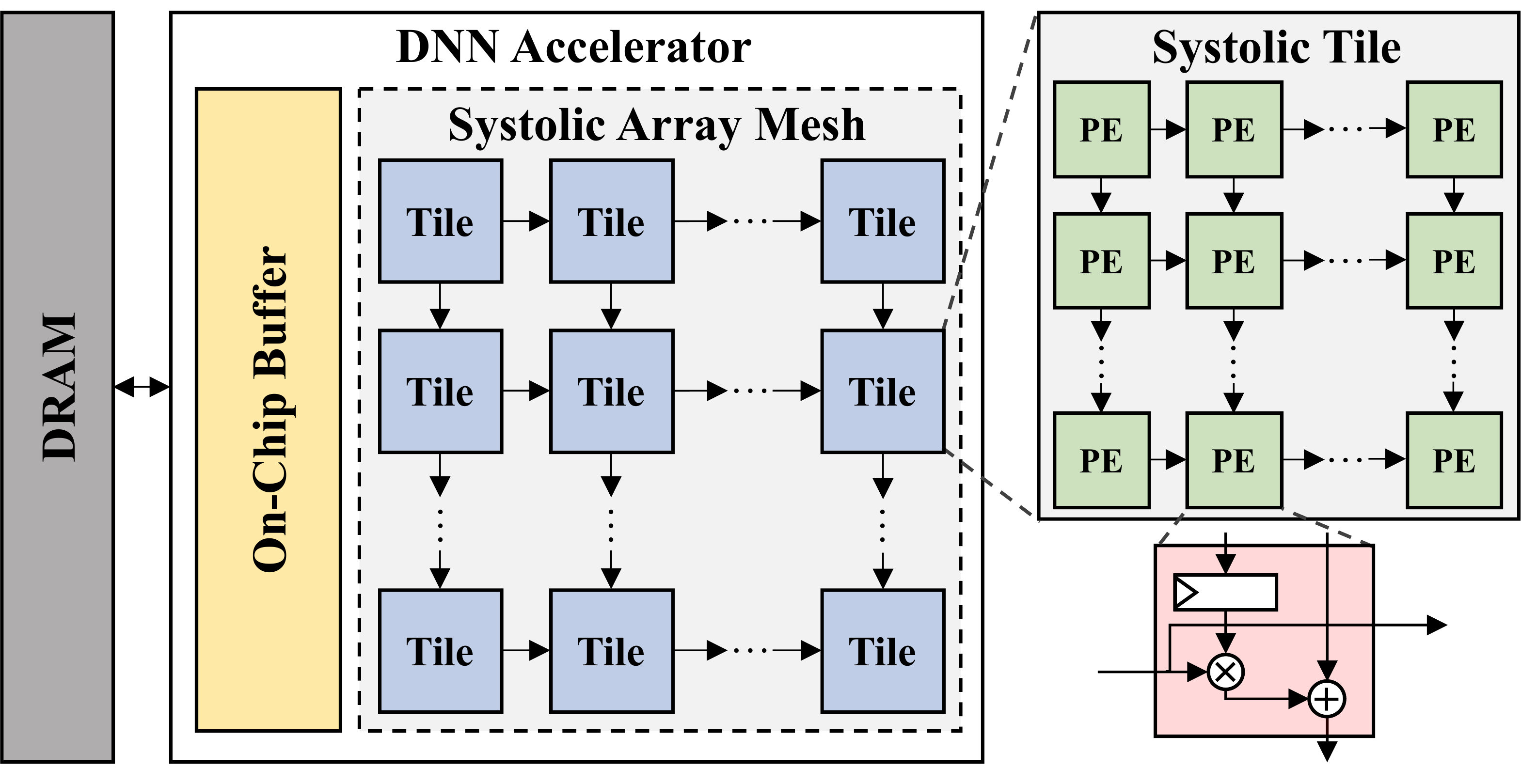}
    \caption{Physical hierarchy of the DNN accelerator architecture.}
    \label{fig:accelerator_arch}
\vspace{-10pt}
\end{figure}

% \xch{Reference Google's ISCA 22, describe this architecture can represent a wide range of classical architectures, including TPU, Eyeriss, Simba, ...}
% In this paper, we focus on a highly parameterizable and general design template for DNN accelerators, emphasizing the optimization of multiplier-accumulator (MAC) array configurations that are closely related to overall PPA metrics~\cite{jouppi2017datacenter}. 
% In this paper, we focus on a highly parameterizable and general design template for DNN accelerators, xxx.
% \TODO{cite Gemmini \& logical.}
In this paper, we focus on the systolic array mesh of a highly parameterizable and general design template for DNN accelerators, inspired by Gemmini~\cite{genc2021gemmini}, emphasizing the optimization of multiplier-accumulator (MAC) array configurations that are closely related to overall PPA metrics~\cite{jouppi2017datacenter}.
As shown in Fig.~\ref{fig:accelerator_arch}, the accelerator architecture comprises a systolic array mesh, integrated with on-chip buffers, and connected to off-chip memory for data storage. 
This mesh is structured as a grid of systolic tiles interconnected with pipeline registers. 
Each systolic tile contains a grid of parallel processing elements (PEs), with each PE executing a single MAC operation. 
By appropriately adjusting the design configurations, the architectural template can emulate a diverse range of representative accelerator architectures, varying in computational power, as well as power and area characteristics.

\subsection{Diffusion Models}

Diffusion models~\cite{sohl2015deep,ho2020denoising} are strong generative models, demonstrating impressive performance in the controllable generation of diverse and high-quality contents~\cite{li2022diffusion,kim2022guided}.
The generation process is formulated as a gradual denoising procedure, beginning with random noise and culminating in the recovery of the original clean data.
Formally, given noisy data obtained by perturbing clean data with random noise in the forward process, the diffusion models are trained to progressively remove noise in a multi-step reverse process, attempting to recover the original clean data.
In the forward process, the diffusion model gradually perturbs initial clean data $x_0$ with Gaussian noise $\epsilon \sim \mathcal{N}(0, \mathbf{I})$:
\begin{equation}
    x_t = \sqrt{\alpha_t} \cdot x_{0} + \sqrt{1 - \alpha_t} \cdot \epsilon,
    \label{eq:diffusion_forward}
\end{equation}
where $x_t$ represents the noisy data at timestep $t = 1, 2, \cdots, T$, and $\{\alpha_t\}_{t=1}^{T}$ denotes the noise schedule.
The diffusion model learns neural network $\epsilon_\theta$ to predict the injected noise $\epsilon$:
\begin{equation}
    \epsilon_\theta(x_t, t) \approx \epsilon = \dfrac{x_t - \sqrt{\alpha_t} \cdot x_{0}}{\sqrt{1 - \alpha_{t}}}
    \label{eq:diffusion_reverse}
\end{equation}

In the reverse process, the diffusion model leverages trained noise predictor $\epsilon_\theta$ to progressively sample less noisy data $x_{t-1}$ from a given noisy input $x_t$.
While vanilla diffusion models typically require $T=1000$ steps for the reverse process, many works have proposed more efficient sampling strategies.
For instance, denoising diffusion implicit models (DDIM)~\cite{song2020denoising} calculate an auxiliary predicted clean data point $\hat{x}_0$ as follows:
\begin{equation}
    \hat{x}_0 = \dfrac{x_t - \sqrt{1 - \alpha_t} \cdot \epsilon_\theta(x_t, t)}{\sqrt{\alpha_t}},
    \label{eq:x0_prediction}
\end{equation}
and deterministically sample $x_{t-1}$ towards $\hat{x}_0$, which produces high-quality samples much faster.

\begin{figure*}[!htbp]
    \centering
    \includegraphics[width=0.7\textwidth]{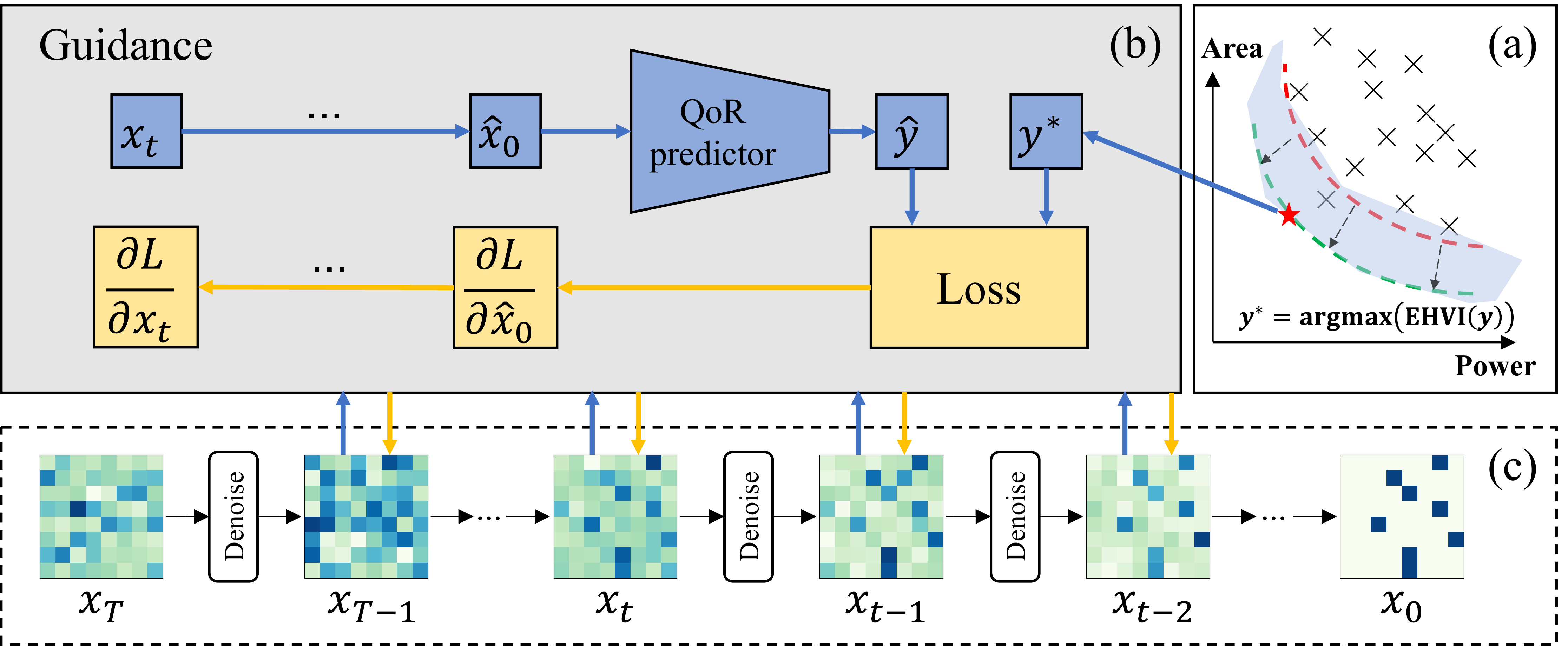}
    \caption{Overview of \papername{} framework. (a) Query module gets the target QoR by maximizing expected hypervolume improvement. (b) guidance module generates a gradient signal given the target QoR. (c) Diffusion module generates the configuration using the gradient signal during the denoising process.}
    \label{fig:overview}
\vspace{-10pt}
\end{figure*}

\subsection{Problem Formulation}

In this paper, our objective is to improve the post-layout Quality-of-Results (QoR) of DNN accelerators, which is affected by both hardware design configurations and EDA synthesis tool parameter settings. 
The MAC array, as a core component of the accelerator, is particularly sensitive to both architectural and tool parameters, making it a central focus of our optimization efforts. 
As shown in TABLE~\ref{tab:design_space}, the tunable options formulate a cross-layer design space, encompassing hardware architecture, logic synthesis, and physical design.

\begin{table}[!t]
\centering
\caption{Cross-Layer Design Space of DNN Accelerators}
\resizebox{\linewidth}{!}{%
\begin{tabular}{|c|c|c|}
\hline
ID & Parameter                                        & Candidate Values             \\
\hline
1  & \texttt{tile\_row}                               & 1,2,4,8,16                   \\
2  & \texttt{tile\_column}                            & 1,2,4,8,16                   \\
3  & \texttt{mesh\_row}                               & 1,2,4,8,16                   \\
4  & \texttt{mesh\_column}                            & 1,2,4,8,16                   \\
\hline
5  & \texttt{target\_clock\_period\_ns}               & 0.2,0.4,0.6,0.8,1.0,1.2,1.4  \\
6  & \texttt{syn\_generic\_effort}                    & none,low,medium,high         \\
7  & \texttt{syn\_map\_effort}                        & none,low,medium,high,express \\
8  & \texttt{syn\_opt\_effort}                        & none,low,medium,high,express,extreme \\
9  & \texttt{auto\_ungroup}                           & true,false                   \\
\hline
10 & \texttt{place\_utilization}                      & 0.3,0.4,0.5,0.6,0.7          \\
11 & \texttt{place\_glo\_max\_density}                & 0.3,0.4,0.5,0.6,0.7          \\
12 & \texttt{place\_glo\_uniform\_density}            & true,false                   \\
13 & \texttt{place\_glo\_cong\_effort}                & auto,low,medium,high \\
14 & \texttt{place\_glo\_timing\_effort}              & medium,high                  \\
15 & \texttt{place\_glo\_auto\_block\_in\_chan}       & none,soft,partial            \\
16 & \texttt{place\_det\_act\_power\_driven}          & true,false                   \\
\hline
\end{tabular}
} %
\label{tab:design_space}
\vspace{-10pt}
\end{table}

\textbf{Definition 1} (Cross-Layer Design Configuration)
\textit{A design configuration is to be defined as a combination of candidate parameter values given in TABLE~\ref{tab:design_space}. }

The parameter combination should satisfy certain constraints.
For instance, the MAC array tile size should not exceed the mesh size, and the maximum global placement density should be no less than floorplan utilization rate.
For a legal design configuration, we assess its QoR from multiple aspects, focusing on both performance and implementation costs related to power and area.

\textbf{Definition 2} (Performance)
\textit{The performance is to be defined as the computational power of the MAC array, which equals the number of MAC units divided by the minimum duration of a clock cycle. }

\textbf{Definition 3} (Power)
\textit{The power is to be defined as the average power dissipation when running benchmark workloads at the maximum attainable design frequency.}

\textbf{Definition 4} (Area)
\textit{The area is to be defined as the size of the floorplan in which the synthesized MAC array is placed.}

Typically, improving the design performance introduces increased implementation cost of power dissipation and area overhead.
To jointly optimize multiple QoR metrics, one eventually arrives at Pareto-optimal design configurations, where one QoR metric cannot be improved without worsening another metric.
To jointly optimize multiple QoR metrics, our framework aims to derive an approximated set of Pareto-optimal solutions with limited trials.
Concretely, the problem can be formulated as follows.

\textbf{Problem 1} (Design Space Exploration) 
\textit{Given the cross-layer design space $\mathcal{D}$, in which a valid design configuration $\mathbf{x}$ can be evaluated for its QoR metrics $\mathbf{y}$ through VLSI flow, the Pareto-driven design space exploration framework aims to identify as many Pareto-optimal design configurations as possible under an upper limit of VLSI flow invocations.}

\section{methodology}
\label{sec:methodology}

\subsection{Framework Overview}

The overview of \papername{} framework is shown in Fig.~\ref{fig:overview}.
% At the core of our framework is a conditional diffusion model, capable of generating diverse design configurations that satisfy design constraints (Section~\ref{sec:design_generation}).
% Given a desired QoR as condition, the diffusion model can conditionally sample promising design configurations meriting further evaluation (Section~\ref{sec:cond_sampling}).
% The selection of conditioning objective value is non-trivial task, as too conservative ones degrades DSE efficiency, while too radical ones deviates far from the realistic prior.
% In \papername{}, we propose Pareto-aware conditioning mechanism to choose proper objective values as conditional prompts, which selects QoR values that reside closely to current Pareto frontier, while maximizing the Expected Hyper-Volume Improvement (EHVI) (Section~\ref{sec:cond_selection}).
The diffusion module is responsible for generating parameter combination akin to training data, aiming to improve the prediction fidelity of cost predictor (Section~\ref{sec:design_generation}).
The guidance module utilizes a target QoR to generate gradient guidance, which steers the diffusion process towards the desired outcomes (Section~\ref{sec:cond_sampling}).
The query module selects proper QoR value as optimization objective, which carefully trades off multiple QoR objectives (Section~\ref{sec:cond_selection}).

\subsection{Diffusion-Based Design Generation}
\label{sec:design_generation}

Generally, the QoR predictor model exhibits higher accuracy for data points akin to those in the training dataset. 
To ensure that the DSE remains within the applicable range of the QoR predictor, we employ diffusion models to generate parameter combinations similar to training data, enabling stable progress towards improved QoR outcomes.
Inspired by the successful application of diffusion models in computer vision~\cite{sohl2015deep,ho2020denoising}, we encode parameter combinations in compact tensors and formulate the diffusion process similarly to image generation.
Specifically, we encode parameter combination as a binary bitmap $x \in \{0, 1\}^{N \times K}$, with $N$ as the total number of parameters, $K$ as the maximum number of candidate values, and $x[i, j] = 1$ indicating the $i$-th parameter is assigned with the $j$-th candidate value.
The diffusion process commences by converting discrete binary bitmap $x$ into a continuous tensor $\tilde{x}$, with each binary bit $b = 0, 1$ mapped to a corresponding real value $r = -1.0, 1.0$.
In the forward process of pretraining, $\tilde{x}$ is first perturbed with random Gaussian noise $\epsilon \sim \mathcal{N}(0, \mathbf{I})$. 
As shown in Fig.~\ref{fig:overview}(c), during the reverse process of inference, the denoising network $\epsilon_\theta$ learns to reconstruct $\tilde{x}$ from noisy states, which is formulated as minimizing the mean-square error between predicted noise $\hat\epsilon$ and $\epsilon$.
The denoised tensor can be quantized back to binary bitmap by decoding each real value to binary bit according to its sign.

Although diffusion models excel at learning complex high-dimensional probability distributions, they may occasionally yield invalid parameter combination.
% To address this issue, We examine possible scenarios of design constraint violations and propose corresponding legalization measures:
To address this issue, we take the following measures:

\begin{itemize}
    \item We examine possible scenarios of design constraint violations and adopt legalization procedure. For instance, if the $i$-th parameter is restricted to be no larger than the $i'$-th parameter, we adjust the $i$-th parameter to the maximum permissible value that suffices the rule.
    \item We employ data augmentation to improve the robustness of diffusion models. By randomly mutating parameter configurations from the original training dataset, we generate new data that helps the diffusion models learn design rules. Notably, the augmented data are unlabeled and do not necessitate additional VLSI evaluations.
\end{itemize}

% \begin{itemize}
%     \item For the $i$-th parameter with $k_i$ candidate values, we choose candidate $j \in [1, K]$ with the largest $x[i,j]$. If $j$ is out of bound, i.e. $j > k_i$, we randomly select a valid candidate value.
%     \item If the $i$-th parameter is restricted to be no larger than the $i'$-th parameter, and current parameter assignment violates this constraint, we adjust the $i$-th parameter to the maximum permissible value that suffices the rule.
% \end{itemize}

\subsection{Conditional Sampling Process}
\label{sec:cond_sampling}

Although the learned diffusion model effectively captures the distribution of the training data, it produces both favorable and unfavorable parameter combinations with comparable probabilities. 
To facilitate efficient exploration, \papername{} employs guidance module to conditionally select promising configurations.
In a nutshell, the guidance module examines the diffusion intermediate state $x_t$, and provides informative feedback to the diffusion process through gradient descent.

During the conditional generation process, the diffusion module and guidance module operate interactively. 
As shown in Fig.~\ref{fig:overview}(b), at each timestep $t$ of the reverse diffusion process, the diffusion module generates the auxiliary predicted clean data point $\hat{x}_0$ (see Equation~\eqref{eq:x0_prediction}), which are then evaluated by the guidance module to predict QoR $\hat{y}=f_\pi(\hat{x}_0)$.
% , where $\pi$ is the weight of QoR predictor. 
Given the target QoR $y^*$, the guidance module computes the loss function $\mathcal{L}(\hat{y}, y^*)$, and calculates gradient signals to indicate how the diffusion module should adjust its trajectory towards closer adherence to the target QoR. To summarize, the following equation gives the refined noise,
\begin{equation}
    \label{eq:guided_noise}
    \hat{\epsilon}_\theta (x_t, t) = \epsilon_\theta (x_t, t) - s(t)\cdot \nabla_{x_t} \mathcal{L}(f_\pi(\hat{x}_0), y^*),
\end{equation}
where $s(t)$ controls the guidance strength.

% To achieve targeted exploration within promising regions of the design space, \papername{} utilizes conditional diffusion models. 
% The model takes normalized design configuration $\tilde{x}$ as input, while the condition information serves as a guiding signal to ensure the generated samples adhere to specific QoR. 
% By leveraging the interplay between the diffusion module and the guidance module, the model achieves conditionally controlled generation of configuration samples.

% The \textbf{diffusion module} is pre-trained on unlabeled configuration data, utilizing a denoising generation framework to iteratively transform noise samples into structured configurations. 
% The training objective is to minimize the denoising error at each step of the diffusion process, enabling the model to capture the underlying distribution of the configuration data, such as constraints and data format. 
% Through this pre-training phase, the diffusion module develops the capacity to model the foundational sample distribution, providing a robust basis for subsequent conditional generation tasks.

The guidance module is trained and retrained using labeled data to capture the relationship between QoR and configuration data. 
During initial training, it leverages the available labeled data to learn how QoR influences the underlying sample distribution, producing gradient signals to guide the diffusion process. 
When new labeled data become available, the guidance module is retrained to accommodate the updated condition distributions. 
The guidance module’s output gradients dynamically steer the diffusion process, ensuring that generated samples adhere to the specified conditions.
This iterative adjustment mechanism ensures that the generated samples not only respect the underlying distribution of the configuration data but also satisfy the constraints imposed by the target QoR, achieving precise and high-quality conditional generation.

\subsection{Pareto-Aware Condition Selection}
\label{sec:cond_selection}

The Pareto-aware condition selection method is designed to identify target QoR values for conditional sampling by leveraging the existing Pareto frontier. As shown in Fig.~\ref{fig:overview}(a), this approach aims to guide the generation process toward optimal configurations by strategically selecting QoR targets that maximize the hypervolume improvement within a defined step size. By focusing on the expansion of the Pareto frontier, this method ensures that the sampling process emphasizes solutions that balance multiple objectives effectively.

The \textit{Pareto frontier} represents the set of non-dominated solutions where improving one QoR objective cannot occur without sacrificing at least one other. Mathematically, for a minimization problem, a point $ x $ is Pareto optimal if no other point $ y $ exists such that $ y_i \geq x_i $ for all $ i $ and $ y_i > x_i $ for at least one $ i $, where $ i $ indexes the objectives. The goal of Pareto-aware condition selection is to maximize the \textit{hypervolume} $ \mathrm{HV}(S) $, defined as the volume of the region dominated by the Pareto set $ S $ and bounded by a reference point $ r $. For a given Pareto set $ S $, the hypervolume is expressed as:
\begin{equation}
    \mathrm{HV}(S) = \int_{r} \mathbf{1}\{\exists s \in S : s \preceq x\} \mathrm{d}x,
\end{equation}
where $ \mathbf{1}\{\cdot\} $ is the indicator function, and $ s \preceq x $ indicates that $ s $ dominates or equals $ x $. Maximizing $ \mathrm{HV} $ ensures that the selected targets contribute to expanding the Pareto frontier into regions of interest.

\begin{figure*}[!htbp]
    \centering
    \includegraphics[width=0.8\textwidth]{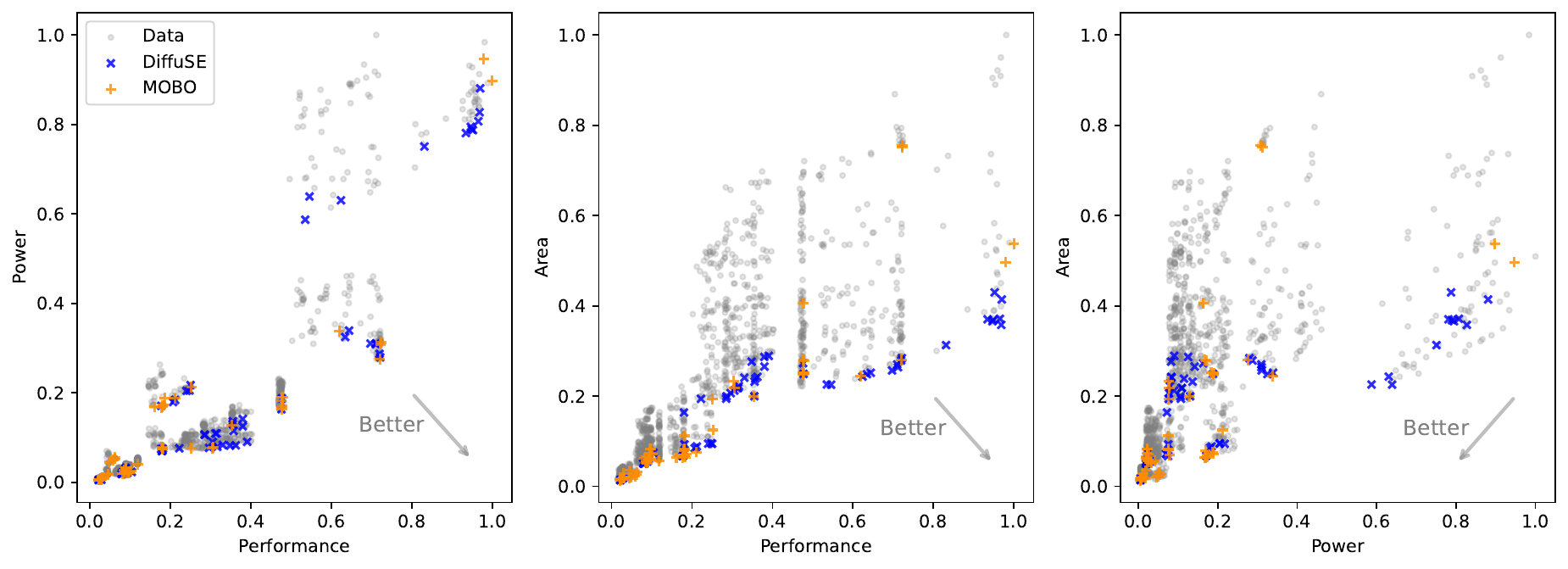}
    \caption{Pareto frontier comparison of normalized QoR between MOBO and \papername{}.}
    \label{fig:pareto}
\vspace{-10pt}
\end{figure*}

To select the next target QoR, the method evaluates candidate points within a predefined step size $\delta$ around the current Pareto frontier. Each candidate point $y$ is assessed for its expected hypervolume improvement $\mathrm{EHVI}(y)$ while adhering to the constraints defined by the step size. The step size ensures a controlled exploration of the objective space, preventing overly aggressive shifts that could destabilize the sampling process. The candidate $y^*$ with the highest hypervolume contribution is chosen as the target QoR for the next sampling iteration, thereby directing the generation process to refine the Pareto frontier iteratively. This selection strategy ensures that the sampling remains aligned with the overarching goal of multi-objective optimization.

\section{experiment}
\label{sec:experiment}

\subsection{Experiment Settings}

\subsubsection{Platform}

The VLSI flow to evaluate QoR for each configuration is run on a Linux-based platform with a Intel(R) Xeon(R) Gold 6342 CPU @ 2.80GHz, and 1536 GiB of memory. All model training and inference are run on a Linux-based platform with an Intel(R) Xeon(R) CPU E5-2698 v4 @ 2.20GHz, 251GiB of memory and an NVIDIA V100 GPU.
Chipyard framework~\cite{amid2020chipyard} is leveraged to compile various Gemmini-based~\cite{genc2021gemmini} systolic array RTL designs. 
We utilize 7-nm ASAP7 PDK~\cite{clark2016asap7} for the VLSI flow. 
Cadence Genus 19.12-s121\_1 and Cadence Innovus v21.14-s109\_1 are used to synthesize and place every sampled RTL design.

\subsubsection{Data Preparation}

The data used in this work consist of both offline and online datasets. The \textit{offline data} includes 10,000 unlabeled data points, randomly sampled from the configuration space, to represent the general distribution of possible configurations. Additionally, a subset of 1,000 data points from the unlabeled set is randomly selected and labeled with the corresponding QoR metrics to form the labeled dataset. This offline dataset provides the foundation for pretraining and initial model tuning. 

The \textit{online data} allows the method to further explore the configuration space beyond the offline dataset. Specifically, the method is permitted to collect up to 256 additional labeled data points during the exploration process. These points are dynamically selected to refine the model's understanding of the QoR distribution, enabling it to more effectively target optimal configurations. This combination of offline and online data ensures a balance between leveraging pre-existing information and exploring new regions of the configuration space for improved performance. 

\subsubsection{Hyperprameter Settings}

% Diffusion module Settings & guidance module settings
For the implementation of diffusion module, we follow the PyTorch version of the previous work~\cite{ho2020denoising}. The QoR predictor used in guidance module is a 3-layer CNN constituted by convolutional residual blocks~\cite{he2016deep}.
% Guidance and self-recurrence
For gradient-guided sampling, we set the Pareto frontier step size $\delta = 0.1$, the guidance strength $s(t) = 1000 \sqrt{1 - \alpha_t}$, DDIM sampling step $S = 50$.

\subsubsection{Baseline}

The baseline for comparison in this study is a multi-objective Bayesian optimization (MOBO) approach, which is widely used in recent DSE approaches~\cite{geng2022ptpt,geng2022ppatuner,bai2021boom}. Specifically, the implementation utilizes Gaussian Process (GP) regression as the surrogate model to approximate the underlying QoR distribution across the configuration space. The acquisition strategy employed is the expected hypervolume improvement (EHVI), which selects new sampling points by maximizing the expected improvement in the Pareto frontier's hypervolume. 

\subsection{Result Analysis}

\subsubsection{Pareto Frontier of QoR}

The Pareto frontier plots in Fig.~\ref{fig:pareto} illustrate the trade-offs between the three objectives: Performance (higher is better), Power, and Area (both lower are better). The proposed \papername{} method is compared against the MOBO baseline across all objective combinations.

In the Performance-Power plot, \papername{} demonstrates a broader and more comprehensive coverage of the Pareto frontier compared to MOBO. It achieves configurations with higher performance while maintaining lower power consumption in several regions, indicating a better exploration of the trade-off space. MOBO, while effective in some areas, appears to converge on fewer high-performance configurations.

Similarly, in the Performance-Area plot, \papername{} consistently finds configurations that outperform MOBO in terms of achieving higher performance with comparable or smaller area requirements. The frontier points obtained by \papername{} extend further toward the desired high-performance and low-area corner, showcasing its ability to balance these two competing objectives effectively.

For the Power-Area plot, the two objectives exhibit a strong positive correlation, where reducing power often leads to a reduction in area, and vice versa. In this context, \papername{}'s advantage lies in exploring a broader range of the power-area space. By covering configurations that range from low power and small area to higher power and larger area, \papername{} provides a more extensive search for optimal trade-offs under different scenarios. In contrast, MOBO produces a more concentrated set of solutions, limiting its ability to address diverse constraints and optimization needs.

Overall, \papername{} demonstrates clear advantages in the performance-power and performance-area trade-offs while showcasing an extensive exploration of the power-area space. This broader coverage ensures that \papername{} can adapt to complex multi-objective optimization tasks, offering diverse and well-distributed solutions across all objective combinations.

\begin{figure}[!tbp]
    \centering
    \includegraphics[width=0.7\linewidth]{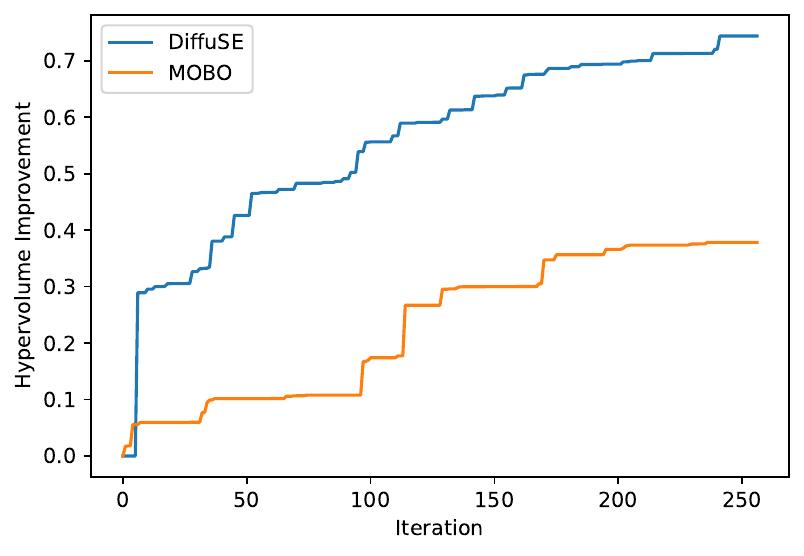}
    \caption{Comparison of the HV improvement between \papername{} and MOBO.}
    \label{fig:hv-comparison}
\vspace{-10pt}
\end{figure}

\subsubsection{Hypervolume of QoR}

Fig.~\ref{fig:hv-comparison} presents the hypervolume improvement (HVI) based on the offline dataset for \papername{} and the MOBO baseline over the course of $256$ online iterations. The plot demonstrates that \papername{} consistently achieves higher hypervolume values compared to MOBO throughout the optimization process. In the early iterations, \papername{} rapidly increases its HV, reflecting its ability to explore the objective space efficiently and identify diverse high-quality solutions. MOBO, on the other hand, shows a slower and less pronounced improvement in HV, indicating a more conservative exploration strategy.

As iterations progress, \papername{} maintains its advantage, with a steady improvement in HV that consistently outperforms MOBO. This suggests that \papername{} not only excels at initial exploration but also continues to refine the Pareto frontier effectively, expanding into regions of the objective space that are challenging for MOBO to reach. By the end of the $256$ iterations, \papername{} achieves an HVI improvement of \hviimprove{} over MOBO, underscoring its superior performance in multi-objective optimization.

% Overall, the hypervolume comparison highlights the effectiveness of \papername{} in achieving broader and more diverse coverage of the objective space compared to the MOBO baseline. This result reinforces \papername{}'s capability to deliver robust and well-balanced solutions in complex optimization scenarios.

% \begin{table}[!t]
% \centering
% \caption{Best Points found by \papername{}}
% \resizebox{\linewidth}{!}{%
% \begin{threeparttable}
% \begin{tabular}{ccccccccc}
% \toprule
% PPA\tnote{\dag} & Perf.\tnote{*} & Timing & Power & Area & Dim & Row & Col & Clock \\
%  ($10^{-5}$) & & ($\mathrm{ps}$) & ($10^{-3} \mathrm{W}$) & ($10^5 \mu\mathrm{m}^2$) & & & & ($\mathrm{ns}$) \\
% \midrule
% 1.19 & 0.662 & 386.8 & 130.6 & 2.83 & 16 & 2 & 8 & 0.4 \\
% 1.17 & 0.333 & 768.9 & 38.7 & 2.44 & 16 & 2 & 2 & 1.4 \\
% 1.24 & 0.085 & 751.7 & 9.7 & 0.60 & 8 & 2 & 8 & 1.4 \\
% 1.14 & 0.165 & 387.7 & 33.0 & 0.72 & 8 & 2 & 2 & 0.4 \\
% 1.48 & 0.026 & 607.0 & 2.6 & 0.18 & 4 & 1 & 4 & 1.4 \\
% 1.24 & 0.020 & 797.6 & 2.3 & 0.14 & 4 & 4 & 2 & 1.4 \\
% \bottomrule
% \end{tabular}
% \begin{tablenotes}
% \footnotesize
% \item[*] $\text{Perf.} = \text{Dim}^2/\text{Timing}$
% \item[\dag] $\text{PPA} = \text{Perf}^2/(\text{Power}\times \text{Area})$
% \end{tablenotes}
% \end{threeparttable}
% }%
% \label{tab:best}
% \end{table}

\subsubsection{Best Points}

TABLE~\ref{tab:best} shows main configurations and QoRs of best points found by \papername{}, and the default point of Gemmini~\cite{genc2021gemmini}.
For each MAC array dimension, the two configurations with the highest PPA trade-off values are listed, where PPA trade-off is defined as $\text{PPA} = \text{Perf}^2/(\text{Power}\times \text{Area})$, following ArchExplorer~\cite{bai2023archexplorer}. 
% \TODO{explain PPA, cite ArchExplorer.}
A clock of $0.4\ \mathrm{ns}$ corresponds to high-performance designs, while $1.4\ \mathrm{ns}$ targets low-power configurations. Larger row and column values indicate larger tiles, which increase power and area but can improve performance in high-performance designs.
% For example, when $\text{Dim} = 16$ and $\text{Clock} = 0.4$, $\text{Row} = 2$ and $\text{Col} = 8$ achieves high performance ($0.662$), while for \textbf{Dim} = 4 and \textbf{Clock} = 1.4, Row = 4 and Col = 2 balances power (2.3 mW) and area (0.14 $\times 10^5 \mu$m\textsuperscript{2}). 
These results show that \papername{} can effectively identify both high-performance and low-cost configurations, demonstrating its flexibility in optimizing chip designs. Compared with Gemmini default configuration, \papername{} achieves \ppaimprove{} improvement in PPA trade-off.

\begin{table}[!t]
\centering
\caption{Gemmini~\cite{genc2021gemmini} Default Point and Best Points Found by \papername{}}
\resizebox{\linewidth}{!}{%
\begin{threeparttable}
\begin{tabular}{ccccccccc}
\toprule
Dim\tnote{*} & Row\tnote{\dag} & Col\tnote{\dag} & Clock & Timing & Power & Area & Perf.\tnote{\ddag} & PPA\tnote{+} \\
 & & & ($\mathrm{ns}$) & ($\mathrm{ps}$) & ($10^{-3} \mathrm{W}$) & ($10^5 \mu\mathrm{m}^2$) & & ($10^{-5}$) \\
\midrule
16 & 1 & 1 & 0.4 & 392.7 & 148.0 & 5.97 & 0.652 & 0.48 \\
\midrule
16 & 2 & 8 & 0.4 & 386.8 & 130.6 & 2.83 & 0.662 & 1.19 \\
16 & 2 & 2 & 1.4 & 768.9 & 38.7  & 2.44 & 0.333 & 1.17 \\
8  & 2 & 8 & 1.4 & 751.7 & 9.7   & 0.60 & 0.085 & 1.24 \\
8  & 2 & 2 & 0.4 & 387.7 & 33.0  & 0.72 & 0.165 & 1.14 \\
4  & 1 & 4 & 1.4 & 607.0 & 2.6   & 0.18 & 0.026 & 1.48 \\
4  & 4 & 2 & 1.4 & 797.6 & 2.3   & 0.14 & 0.020 & 1.24 \\
\bottomrule
\end{tabular}
\begin{tablenotes}
\footnotesize
\item[*] $\text{Dim} = \text{TileRow}\times\text{MeshRow} = \text{TileCol}\times\text{MeshCol}$
\item[\dag] Row and Col refer to TileRow and TileCol, respectively.
\item[\ddag] $\text{Perf.} = \text{Dim}^2/\text{Timing}$
\item[+] $\text{PPA} = \text{Perf}^2/(\text{Power}\times \text{Area})$
\end{tablenotes}
\end{threeparttable}
}%
\label{tab:best}

\vspace{-10pt}
\end{table}

% For each dimension, the table lists the two configurations with the highest PPA tradeoff values. The configurations illustrate a clear relationship between clock settings and the nature of the chip design: a clock of 0.4 ns corresponds to high-performance designs, while a clock of 1.4 ns aligns with low-power, low-cost configurations. 

% Within each dimension and clock setting, the configurations with larger row and column values correspond to larger tile sizes, as increasing row and column dimensions results in larger pure logic circuits. This is evident in the higher power and area values for such configurations, which may lead to higher performance in high-performance designs but are less optimal for low-power configurations. For instance, for \textbf{Dim} = 16 and \textbf{Clock} = 0.4, the configuration with Row = 2 and Col = 8 achieves high performance (0.662) due to its larger tile size, while for \textbf{Dim} = 4 and \textbf{Clock} = 1.4, Row = 4 and Col = 2 achieves a balance with lower power (2.3 mW) and area (0.14 $\times 10^5 \mu$m\textsuperscript{2}).

% Overall, the results demonstrate that \papername{} effectively identifies both high-performance designs and low-cost configurations, showcasing its versatility and potential for discovering promising chip designs under varying constraints.

\begin{table}[!t]
\centering
\caption{Hyperparameter Sensitivity}
\resizebox{0.8\linewidth}{!}{%
\begin{tabular}{cc|cc}
\toprule
Step Size       & Guidance Strength & HV Improvement    & Error Rate        \\ 
\midrule
0.05            & 1000              & 0.516             & \textbf{3.9\%}    \\ 
\textbf{0.10}   & \textbf{1000}     & \textbf{0.744}    & 4.7\%             \\ 
0.10            & 2000              & 0.431             & 15.2\%            \\ 
\bottomrule
\end{tabular}
}%
\label{tab:hyperparameter}
\vspace{-15pt}
\end{table}

\subsection{Hyperparameter Sensitivity}

TABLE~\ref{tab:hyperparameter} presents the sensitivity analysis of the key hyperparameters: step size and guidance strength, with respect to the resulting HV improvement (based on the 1000 offline data) and configuration error rate. 
Increasing guidance strength will give stronger guidance to the diffusion module, but may disrupt the natural generation process. 
Similarly, larger step size will increase the range of QoR exploration, but risk deviating from the data distribution.
% The analysis aims to identify the hyperparameter configuration that maximizes optimization performance while maintaining acceptable accuracy.
The best result, highlighted in bold, is achieved with a step size of 0.10 and a guidance strength of 1000. This configuration produces the highest HV improvement of 0.744 while maintaining a relatively low error rate of 4.7\%. 
This configuration balances the influence of the guidance module and the diffusion module effectively. 

% The best result, highlighted in bold, is achieved with a step size of 0.10 and guidance strength of 1000. This configuration produces the highest HV improvement of 0.744 while maintaining a relatively low error rate of 4.7\%. As a result, this configuration is selected as the default setting for the proposed method.

% Increasing guidance strength will disrupts the natural generation process of the diffusion module, leading to instability and poor-quality samples. Conversely, overly low Guidance Strength diminishes the effectiveness of the guidance module, resulting in insufficient alignment with the target conditions. The chosen value of 1000 strikes a balance, providing effective guidance without compromising the generation quality.

% The sensitivity analysis emphasizes the importance of carefully tuning hyperparameters to achieve a balance between HV improvement and error rate. The chosen configuration provides an effective trade-off, demonstrating robust performance in multi-objective optimization scenarios.

\section{conclusion}
\label{sec:conclusion}

We proposed \papername{}, a diffusion-driven framework for exploring cross-layer DNN accelerator design spaces. By leveraging conditional diffusion models and Pareto-aware conditioning, \papername{} achieves superior trade-offs among performance, power, and area. 
% Compared to the Gemmini default configuration, DiffuSE achieves significant improvements, including up to a \ppaimprove{} increase in PPA trade-off. 
% Experimental results show that \papername{} outperforms the MOBO baseline in Pareto frontier coverage and hypervolume improvement, effectively identifying both high-performance and low-cost designs. 
% These results highlight \papername{}'s efficiency and scalability in optimizing complex design spaces, offering a promising approach for automated hardware design.
Experiments show that \papername{} outperforms the MOBO baseline in Pareto frontier coverage and hypervolume improvement, highlighting \papername{}'s efficiency and scalability in optimizing complex cross-layer design spaces.
In future work, we plan to extend this work by optimizing the memory hierarchy of DNN accelerators and incorporating additional physical design parameters at clock tree synthesis and routing stages.

\section*{Acknowledgement}
This work is supported in part by Beijing Natural Science Foundation (Grant No. L243001), National Natural Science Foundation of China (Grant No. 62032001), and 111 Project (B18001).

% \clearpage

{
\bibliographystyle{IEEEtran}
\bibliography{./references}
}

\end{document}